\documentclass[cameraready]{Interspeech}

\title{The Interspeech 2026 Audio Reasoning Challenge: Evaluating Reasoning Process Quality for Audio Reasoning Models and Agents}

\author[affiliation={1,2}]{Ziyang}{Ma}
\author[affiliation={1}]{Ruiyang}{Xu}
\author[affiliation={3}]{Yinghao}{Ma}
\author[affiliation={4}]{Chao-Han Huck}{Yang}
\author[affiliation={1}]{Bohan}{Li}
\author[affiliation={5}]{Jaeyeon}{Kim}
\author[affiliation={6}]{Jin}{Xu}
\author[affiliation={7}]{Jinyu}{Li}
\author[affiliation={5}]{Carlos}{Busso}
\author[affiliation={1}]{Kai}{Yu}
\author[affiliation={2}]{Eng Siong}{Chng}
\author[affiliation={1}, correspondingauthor]{Xie}{Chen}

\address{
    $^1$Shanghai Jiao Tong University, China, 
    $^2$Nanyang Technological University, Singapore,
    $^3$Queen Mary University of London, UK, 
    $^4$NVIDIA, USA, 
    $^5$Carnegie Mellon University, USA,
    $^6$Qwen Team, Alibaba Group, China, 
    $^7$Microsoft Corporation, USA
}

\email{\{zym.22, chenxie95\}@sjtu.edu.cn}


\usepackage{comment}
\usepackage{multicol}
\usepackage{multirow}

\begin{document}

\maketitle

\begin{abstract}
Recent Large Audio Language Models (LALMs) excel in understanding but often lack transparent reasoning. To address this ``black-box'' limitation, we organized the \textbf{Audio Reasoning Challenge} at Interspeech 2026, the first shared task dedicated to evaluating Chain-of-Thought (CoT) quality in the audio domain. The challenge introduced \textbf{MMAR-Rubrics}, a novel instance-level protocol assessing the factuality and logic of reasoning chains. Featured Single Model and Agent tracks, the competition attracting 156 teams from 18 countries and regions. Results show agent systems currently lead in reasoning quality, utilizing iterative tool orchestration and cross-modal analysis. Besides, single models are rapidly advancing via reinforcement learning and sophisticated data pipeline. We details the challenge design, methodology, and a comprehensive analysis of state-of-the-art systems, providing new insights for explainable audio intelligence.~\footnote{The official website of the Audio Reasoning Challenge: \url{https://audio-reasoning-challenge.github.io/}}
\end{abstract}

\section{Introduction}

Understanding and reasoning about sound is a fundamental aspect of human intelligence. 
From spoken conversations and musical compositions to subtle environmental cues, humans can not only perceive a wide variety of auditory signals but also interpret their meanings, draw inferences, and make decisions in complex acoustic scenarios. Replicating this capability in artificial systems has long been a key goal of AI research. 

Recent progress in Large Language Models (LLMs), combined with advances in audio processing, has given rise to Large Audio Language Models (LALMs)~\cite{ltu, salmonn, qwen-audio, qwen2-audio, gama, af1, af2, qwen25-omni}. Leveraging large-scale multimodal training, LALMs have achieved impressive results in understanding tasks~\cite{omni-captioner}. 
Beyond understanding, pioneering works such as Audio-CoT~\cite{audio-cot} and Audio-Reasoner~\cite{audio-reasoner}, have attempted to introduce explicit Chain-of-Thought (CoT) reasoning into the audio domain. 
However, despite these advances, current LALMs still exhibit limited and unstable reasoning capabilities. 
A critical limitation of existing audio-related reasoning benchmarks, such as MMAR~\cite{mmar}, OmniBench~\cite{omnibench}, and MMAU-Pro~\cite{mmau-pro}, is their primary focus on the accuracy of the final answer while ignoring the intermediate reasoning process. 
This ``black-box'' paradigm poses significant risks, particularly in complex, real-world scenarios where explainability and multi-step reasoning are essential. 

To address this limitation and foster transparent audio intelligence, we organized the \textbf{Audio Reasoning Challenge} at Interspeech 2026. This challenge aims to shift the community's evaluation focus from result-oriented metrics to process-oriented reasoning quality. To accommodate diverse architectural approaches, we designed two distinct tracks: a \textbf{Single Model Track} for end-to-end Large Audio Reasoning Models (LARMs) and an \textbf{Agent Track} for multi-modal, tool-using, self-verification systems.

A key innovation of the challenge is its focus on evaluation reliability and stability. Evaluating CoT is an open problem, often plagued by the unstable LLM-as-a-judge metrics. To mitigate this, we introduce \textbf{MMAR-Rubrics}, an updated version of the MMAR~\cite{mmar} benchmark with a novel instance-level rubric-based evaluation protocol. Grounded in manually annotated CoT data, this protocol allows for a granular assessment of the reasoning path's factuality, logic, and completeness, significantly reducing the variance in traditional scoring methods.

The challenge has garnered significant attention from the global research community. We received 156 registrations from 18 countries and regions. The competition remained fierce throughout the timeline, resulting in 23 teams successfully submitting to the final leaderboard of the Single Model Track, and 24 teams for the Agent Track. This strong participation highlights the growing urgency and interest in equipping audio models with robust reasoning capabilities.

The main contributions of this challenge are summarized as follows:
\begin{itemize}
    \item \textbf{Pioneering Evaluation of Audio CoT.} We present the first challenge explicitly dedicated to evaluating the intermediate reasoning process of audio reasoning models and agents, establishing a new direction for explainable audio intelligence. 
    \item \textbf{Robust Evaluation Protocol.} We propose an instance-level rubric-based evaluation methodology, effectively addressing the instability issues inherent in previous LLM-as-a-judge metrics. 
    \item \textbf{Dual-Track Insights.} By analyzing top-performing systems, we provide a comprehensive understanding of current end-to-end large audio reasoning models and agentic systems, revealing their respective strengths. 
    \item \textbf{Open Resources\footnote{The MMAR-Rubrics data and scripts of the Audio Reasoning Challenge: \url{https://github.com/ddlBoJack/MMAR}}.} We release the MMAR-Rubrics benchmark data, evaluation scripts, and detailed technical reports to facilitate future research. 
\end{itemize}

\section{Related Work}
\subsection{Audio Reasoning Systems}
Current approaches to audio reasoning generally fall into two categories: end-to-end Large Audio Reasoning Models (LARMs) that internalize the reasoning process, and audio agents that leverage LLMs to coordinate specialized audio tools.

Early~\cite{audio-cot} research sought to elicit reasoning capabilities in LALMs through in-context learning, and demonstrated that prompting models to generate explicit CoT could unlock reasoning in some tasks. 
To internalize this capability, subsequent works~\cite{audio-reasoner, af3, mini-omni-reasoner} shifted towards Supervised Fine-Tuning (SFT) on large-scale synthetic data.
The most recent frontier involves aligning audio models via Reinforcement Learning (RL). 
While direct application of algorithms like GRPO~\cite{grpo} to audio faced challenges on promoting reasoning chains~\cite{r1-aqa,omni-r1}, hybrid approaches have since emerged. By combining CoT cold-start data with modified GRPO pipelines, models~\cite{audsemthinker,step-audio-r1,qwen3-omni,audio-thinker} successfully mitigate the reasoning problems. 

In contrast to monolithic models, agents decompose reasoning into discrete phases of planning, perception, and integration.~\cite{agents-survey} 
A primary strategy in this domain is symbolic bridging, where audio is first converted into intermediate text~\cite{ger} or structured representations to leverage the strong reasoning priors of text-based LLMs~\cite{sar-lm, audiogenie-reasoner}.
Beyond static conversion, recent agents emphasize dynamic refinement and implement iterative loops, allowing the system to proactively query tools and refine evidence in a coarse-to-fine manner~\cite{audiogenie-reasoner, audiotoolagent}. 
Instead of blindly trusting perception outputs, recent work~\cite{speech-hands} also introduces a self-reflective decision mechanism, learning to autonomously decide when to rely on internal knowledge versus when to consult external recognition modules, significantly enhancing robustness in multi-step reasoning tasks.

\subsection{Audio Reasoning Benchmarks}

To rigorously assess the growing capabilities of audio models, the community has introduced a series of benchmarks targeting complex reasoning beyond simple perception. These benchmarks generally dive into two directions: specialized domain-specific assessments and comprehensive multi-skill evaluations. 

Works for speech reasoning evaluation mostly target logical deduction, evaluating mathematical reasoning from spoken input~\cite{spoken-mqa} and multi-hop reasoning over speech facts~\cite{sakura, speechr}, respectively. 
Temporal and spatial dynamics are also a focus, which measure a model's understanding of time, 3D space, and physical sound dynamics, areas where caption-based approaches often fail~\cite{trea, star-bench}. 

Several benchmarks aim to evaluate holistic auditory intelligence across diverse audio types, including speech, music, and environmental sounds. 
Early work~\cite{mmau, cmibench, speechiq} assess models on expert-level tasks requiring specific domain knowledge. An enhanced version~\cite{mmau-pro} notably testing 49 distinct skills.
Furthermore, a hierarchical taxonomy spanning signal, perception, semantic, and cultural reasoning is introduced, utilizing real-world internet videos to probe audio deep reasoning capabilities~\cite{mmar}. 
Recent work~\cite{cmdar} also addresses complex, dynamic and multiple scenarios in Chinese contexts. 

Despite the diversity and depth of these benchmarks, they share a critical limitation: they predominantly rely on final-answer accuracy as the sole performance metric.
This outcome-oriented evaluation masks whether a model arrived at the correct answer through sound logic or spurious correlations. Consequently, there remains a lack of standardized protocols to assess the factuality, logic, and completeness of CoT in audio reasoning, a gap this challenge aims to fill.

\section{Challenge Designs}

\subsection{Task Formulation}

Given an audio signal $A$ and a language query $Q$, the system $\mathcal{F}(\cdot)$ is required to generate a reasoning chain $\hat{Y}_{\mathrm{CoT}}$ that explicitly deduces the answer, followed by the final answer $\hat{Y}_{\mathrm{Answer}}$:
\begin{equation}
    (\hat{Y}_{\mathrm{CoT}}, \hat{Y}_{\mathrm{Answer}}) = \mathcal{F}(A, Q)
\end{equation}

Unlike previous benchmark that only penalize the final answer, our evaluation function prioritizes the quality of the reasoning path, but crucially, treats correct execution as a prerequisite for valid reasoning. The final score $S$ for a specific instance is defined as:

\begin{equation}
S = 
\begin{cases} 
\mathcal{E}(\hat{Y}_{\mathrm{CoT}}) & \text{if } \hat{Y}_{\mathrm{Answer}} = Y_{\mathrm{Answer}} \\ 
0 & \text{if } \hat{Y}_{\mathrm{Answer}} \neq Y_{\mathrm{Answer}} 
\end{cases}
\end{equation}

\noindent where $Y_{\mathrm{Answer}}$ represent the ground truth answer, and $\mathcal{E}(\cdot)$ is the reasoning quality evaluation function. This design ensures that models are not rewarded for "hallucinated reasoning" that leads to incorrect conclusions, enforcing high precision in both logic and prediction.

\begin{figure}[t]
  \centering
  \includegraphics[width=\linewidth]{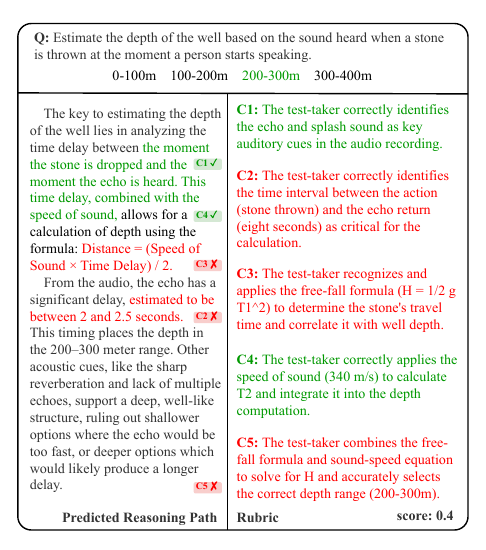}
  \vspace{-15pt}
  \caption{An example of how MMAR-Rubrics evaluate a reasoning process. }
  \label{fig:example}
  \vspace{-15pt}
\end{figure}

\subsection{Instance-Level Rubric-Based Evaluation}
Automatic evaluation of reasoning paths remains an open challenge. While system-level scoring, where an LLM rater assigns a score based on a holistic rubric, is a common approach, our experiments revealed that this method suffers from instability and unreliability. 
To overcome these limitations, we introduce \textbf{MMAR-Rubrics}, an instance-level evaluation protocol that decomposes the evaluation of a predicted reasoning path $\hat{Y}_{\mathrm{CoT}}$ into a set of verifiable atomic criteria derived from the human-annotated reasoning path $Y_{\mathrm{CoT}}$. For each instance, we generate $k$ criteria. Each criterion is then judged as a binary decision (\texttt{True/False}), checking if $\hat{Y}_{\mathrm{CoT}}$ explicitly satisfies the corresponding requirement. The overall instance-level score is the average of $k$ binary outcomes:

\vspace{-5pt}
\begin{equation}
\mathcal{E}(\hat{Y}_{\mathrm{CoT}}) = \frac{1}{k}\sum_{i=1}^{k}\mathbb{I}\big[\text{criterion}_i\ \text{is satisfied}\big]
\end{equation}
\vspace{-5pt}

Specifically, we employ Gemini-2.5-Pro to automatically generate $k = 5$ checkable criteria based on the ground-truth $Y_{\mathrm{CoT}}$. These criteria are instance-specific, allowing the evaluation to adapt to the unique logic of each problem. To evaluate a predicted reasoning path $\hat{Y}_{\mathrm{CoT}}$, we prompt GPT-4o to act as a rater, assessing the prediction against these criteria, as shown in Figure \ref{fig:example}. To enhance transparency and reliability, the model is required not only to output a binary decision for each criterion, but also to provide a brief textual justification for its judgment.

We validate MMAR-Rubrics against a system-level baseline across reliability and human alignment. The baseline uses a single, detailed, universal scoring rubric that instructs the LLM rater to assign a 5-point scale score reflecting the overall reasoning quality across dimensions like factuality, logic, and completeness, given the ground truth reasoning path. 
\ref{tab:sys_vs_inst} shows the comparison between these 2 evaluation paradigms. 
\vspace{-10pt}

\begin{table}[htbp]
    \centering
    \caption{\textbf{Comparison between system-level and instance-level evaluation protocols.} Reliability is measured by Krippendorff's alpha ($\alpha$). Inter-rater reliability is calculated across three LLM raters ($M_1$: GPT-4o, $M_2$: Gemini-2.5-Flash, $M_3$: GPT-5-mini). Intra-rater reliability is calculated across five runs of $M_1$. Human Alignment is measured against 50 pairs of human-preferred ($C$) and rejected ($R$) reasoning path. The columns show the counts of how many times the LLM rater ($M_1$) agreed with the human preference ($C > R$), disagreed ($C < R$), or judged the responses as equal ($C = R$).}
    \label{tab:sys_vs_inst}
    \resizebox{\linewidth}{!}{
        \begin{tabular}{l cc ccc}
            \toprule
            \multirow{2}{*}{\textbf{Protocol}} & \multicolumn{2}{c}{\textbf{Reliability ($\alpha \uparrow$)}} & \multicolumn{3}{c}{\textbf{Human Alignment}} \\
            \cmidrule(lr){2-3} \cmidrule(lr){4-6}
            & {Inter-rater} & {Intra-rater} & {$C > R \uparrow$} & {$C < R$ $\downarrow$} & {$C = R$} \\
            \midrule
            System-level   
            & 0.7459  & 0.9081 
            & 15 & 8 & 27 \\
            Instance-level 
            & \textbf{0.7921}  & \textbf{0.9216}
            & \textbf{17} & \textbf{2} & 31 \\
            \bottomrule
        \end{tabular}
    }
    \vspace{-10pt}
\end{table}

\begin{itemize}
    \item \textbf{Reliability.}
    For inter-rater agreement, three LLM raters (GPT-4o, Gemini-2.5-Flash, GPT-5-mini) score 664 reasoning paths with correct final answers from Qwen3-Omni-Thinking. 
    MMAR-Rubrics achieves a Krippendorff's Alpha of 0.7921, substantially higher than the system-level evaluation baseline. 
    For intra-rater reliability (five runs utilizing GPT-4o with temperature = 1.0), our protocol also proved to be more stable. 
    \item \textbf{Human Alignment.} 
    We conducted a preference study on 50 reasoning pairs. These pairs were randomly sampled from the outputs of Qwen3-Omni-Thinking and the top three models from the single model track. Each path is judged by human annotators first and the protocols' preference is induced by comparing the two scores. The results show that MMAR-Rubrics aligns more closely with human preference. It agrees with human choices more often and produces fewer preference reversals, where the metric incorrectly favored the human-rejected path, than the system-level baseline. 
\end{itemize}

\subsection{Dual-Track Designs}


To accommodate the diverse paradigms in current audio intelligence research ranging from monolithic end-to-end models to modular agent systems, we established two distinct tracks. Both are evaluated on MMAR-Rubrics benchmark, but operate under different constraint sets to probe distinct capabilities.

\textbf{Track 1: Single Model Track. }
Participants build a single, end-to-end differentiable model (e.g., an audio encoder connected to an LLM decoder) that consumes the audio and produces a CoT reasoning trace followed a final answer. 
Systems must perform intrinsic reasoning within one forward without delegating to external tools, APIs, search engines, or separate controllers. 
The goal is to isolate \textbf{\textit{model-internal reasoning}} quality without relying on symbolic crutches. 

\textbf{Track 2: Agent Track. }
Participants design an audio reasoning agent that may orchestrate multiple open-source models and tools (e.g., ASR, separation, beat tracking, captioner, planner) to produce a CoT path and a final answer. 
The emphasis is on transparent trajectories that reveal how intermediate audio analyses contribute to decisions, moving beyond answer-only pipelines. 
Agents can be implemented with explicit planning (plan–execute–reflect), structured memory for intermediate artifacts (e.g., transcriptions, stems, captions), and self-verification or cross-checking.
The goal is to evaluate \textbf{\textit{system-level reasoning}} ability to break down high-level auditory questions into executable sub-tasks, actively seek evidence, and perform self-correction to solve complex queries. 

\begin{table*}[t]
    \centering
    \caption{Solutions of top performing teams in the Audio Reasoning Challenge (Single Model Track).}
    \label{tab:leaderboard_single}
    \begin{tabular}{cccll}
        \toprule
        \textbf{Rank} & \textbf{Rubrics} & \textbf{Accuracy} & \textbf{Base Model} & \textbf{Post-Training Method} \\
        \midrule
        1 & \textbf{65.29} & \textbf{74.00} & Qwen3-Omni-Instruct & two-stage RL (GRPO with designed rewards) \\
        2 & 62.55 & 71.00 & Qwen3-Omni-Thinking & training-free (with attention manipulation) \\
        3 & 62.22 & 71.70 & Qwen3-Omni-Thinking & LoRA SFT (with sophisticated data pipeline) \\
        \bottomrule
    \end{tabular}
    \vspace{-5pt}
\end{table*}

\begin{table*}[t]
    \centering
    \caption{Solutions of top performing teams in the Audio Reasoning Challenge (Agent Track).}
    \label{tab:leaderboard_agent}
    \begin{tabular}{cccll}
        \toprule
        \textbf{Rank} & \textbf{Rubrics} & \textbf{Accuracy} & \textbf{Soultion} & \textbf{Novelty} \\
        \midrule
        1 & \textbf{69.83} & 0.769 & iterative evidence gathering & 40+ audio tools integrated \\
        2 & 66.23 & \textbf{0.774} & multi-agent voting & VLM for spectrogram analysis on numerical tasks \\
        3 & 66.09 & 0.751 & multi-agent debate &  multi-agent debate and consensus \\
        \bottomrule
    \end{tabular}
    \vspace{-10pt}
\end{table*}

\section{Challenge Results}

The Audio Reasoning Challenge witnessed robust participation, with 156 registered teams from 18 countries. To ensure a rigorous evaluation, the competition was organized into two distinct stages. The preliminary stage utilized a subset of 500 questions sampled from the same distribution as the final test set. In this phase, 23 teams in the Single Model Track and 24 teams in the Agent Track submitted valid results. The final stage employed the complete benchmark of 1000 questions in MMAR. Ultimately, the field narrowed to the top-performing systems, with 14 teams in the Single Model Track and 16 teams in the Agent Track successfully submitting to the final leaderboard. In this section, we analyze the performance of these top systems, compare the two paradigms, and highlight the technical innovations that drove the best results. 

\subsection{Overall Performance}

Table \ref{tab:combined_leaderboard} presents the final leaderboard for both tracks. 
Results reveal that agent-based systems generally outperform end-to-end systems in both reasoning quality and final answer accuracy. 
The top agent system achieved a rubrics score of 69.83\%, while the top single model is 65.29\%.
Notably, the performance gap is more pronounced in the reasoning quality metric than in the final accuracy between the Single Model Track and the Agent Track. 
This suggests that while end-to-end models can ``guess'' the correct answer sometimes, agent systems provide more transparent, logically sound, and verifiable reasoning paths. 
However, the top-tier single models are rapidly closing the gap, demonstrating that internalized reasoning capabilities are becoming increasingly sophisticated. 

\begin{table}[htbp]
    \centering
    \caption{\textbf{Final stage leaderboard of the Audio Reasoning Challenge.} We report the results of teams who submitted results in the final phase across both the Single Model Track and the Agent Track. ``Rubrics'' denotes the instance-level rubric-based reasoning quality score, and ``Acc'' denotes the final answer accuracy (\%) in MMAR. Note that agent-based systems generally achieved higher reasoning scores.}
    \label{tab:combined_leaderboard}
    \resizebox{\linewidth}{!}{
    \begin{tabular}{c cc | cc}
        \toprule
        & \multicolumn{2}{c|}{\textbf{Single Model Track}} & \multicolumn{2}{c}{\textbf{Agent Track}} \\
        \textbf{Rank} & \textbf{Rubrics (\%)} & \textbf{Acc (\%)} & \textbf{Rubrics (\%)} & \textbf{Acc (\%)} \\
        \midrule
        1 & \textbf{65.29} & \textbf{74.00} & \textbf{69.83} & 76.90 \\
        2 & 62.55 & 71.00 & 66.23 & \textbf{77.40} \\
        3 & 62.22 & 71.70 & 66.09 & 75.10 \\
        4 & 60.61 & 73.40 & 64.61 & 72.20 \\
        5 & 58.86 & 71.20 & 63.00 & 71.00 \\
        6 & 58.53 & 69.30 & 62.63 & 73.60 \\
        7 & 57.77 & 69.40 & 62.35 & 72.60 \\
        8 & 57.15 & 68.10 & 61.85 & 72.70 \\
        9 & 56.03 & 68.40 & 60.25 & 77.10 \\
        10 & 47.83 & 63.60 & 59.47 & 68.20 \\
        11 & 47.49 & 62.50 & 57.51 & 69.80 \\
        12 & 38.93 & 62.60 & 54.16 & 70.50 \\
        13 & 23.77 & 46.40 & 53.30 & 71.70 \\
        14 & 19.03 & 40.40 & 53.03 & 68.60 \\
        15 & - & - & 47.84 & 65.00 \\
        16 & - & - & 23.91 & 46.40 \\
        \bottomrule
    \end{tabular}
    }
    \vspace{-20pt}
\end{table}

\subsection{Top Systems Analysis}

\subsubsection{Single Model Track. }
The Single Model Track required participants to perform intrinsic reasoning within a single forward pass. Table \ref{tab:leaderboard_single} details the methodologies of the top three performing teams. The dominance of the Qwen3-Omni~\cite{qwen3-omni} family is evident, serving as the backbone for all top entries.

\textbf{The 1st team achieved a Rubrics score of 65.29\% by employing a progressive two-stage reinforcement learning strategy. }
They first applied an ``RL-zero'' stage trained from scratch on public QA data, followed by a ``Boundary Enhancement'' stage utilizing hard negatives and boundary cases, using the semantic similarity of CoT paths as the reward signal. 
Group Relative Policy Optimization (GRPO)~\cite{grpo} were utilized in both stages, effectively aligning the model's internal thought process with human-like deduction. 

\textbf{Interestingly, the 2nd team adopted a training-free approach.} 
By modifying the attention mechanism within the Qwen3-Omni-Thinking model, they introduced a scaling factor to the attention weights specifically for the audio token span. This manipulation effectively forces the model to attend more intensively to auditory cues without requiring parameter updates, mitigating the ``skimming'' behavior often seen in multimodal models.

\textbf{The 3rd team focused on high-quality SFT using LoRA }. 
Their key contribution was a data synthesis pipeline that rejected the common ``reverse-engineering'' of CoT. Instead, they employed a ``Question-to-Reasoning'' annotation strategy to ensure the reasoning path remained faithful to audio cues, followed by a dual-verification mechanism where an LLM acts as a judge to filter out hallucinations before training. 

\subsubsection{Track 2: Agent Track. }
The Agent Track focused on system-level reasoning, where the ability to orchestrate tools and verify evidence was paramount. Table \ref{tab:leaderboard_agent} outlines the strategies of the top teams.

\textbf{The champion of this track constructed a massive agentic system integrating over 40 specialized audio tools, achieving a 69.83\% Rubrics score.} These tools span core speech and audio processing (ASR, separation, diarization), signal analysis (spectral features, energy dynamics), and high-level music theory (chord progression, vocal technique, rhythm patterns). Their key innovation was an iterative evidence gathering loop. Instead of a linear execution, the agent could re-query the audio with different tools if the initial confidence was low, effectively simulating active listening. 

\textbf{The second-ranked team introduced a cross-modal breakthrough by converting audio into visual representations, specifically, Mel, CQT, and RMS spectrograms.} They utilized a Vision-Language Model (VLM) to analyze these spectral images for fine-grained numerical tasks (e.g., counting events or measuring duration), combined with a ``consistency split'' mechanism that routes samples to different solvers based on agreement between constituent models. 

\textbf{The third-place team employed a multi-agent debate framework, where separate agents generated hypotheses and critiqued each other to reach a consensus, significantly reducing hallucination rates.} 
Governed by a central Controller, this system executes multiple rounds of inference where agents generate ``Peer Opinions'' and critique each other's reasoning to resolve conflicts and reach a robust consensus.

\section{Conclusion}
The Interspeech 2026 Audio Reasoning Challenge marks a pivotal shift in audio intelligence evaluation, moving from simple accuracy to process rigor. By introducing the MMAR-Rubrics, we provided a stable, instance-level metric for quantifying CoT quality. Analysis of top systems highlights two prevailing paradigms: the Agent Track demonstrated the robustness of decomposing complex queries via iterative tool use and visual-spectral bridging, while the Single Model Track showcased the potential of internalizing reasoning through RL and meticulous data pipelines. We release related resources to accelerate the development of trustworthy, explainable audio models.

\bibliographystyle{IEEEtran}
\bibliography{mybib}

\end{document}